\documentclass[amssymb,prd,aps]{revtex4}
\usepackage{amssymb}
\usepackage[]{psfig}

\usepackage{latexsym}
\newcommand{\be}{\begin{equation}}
\newcommand{\ee}{\end{equation}}
\newcommand{\ba}{\begin{eqnarray}}
\newcommand{\ea}{\end{eqnarray}}

\begin{document}

\draft
\title{Dyons in ${\cal N}=4$ Gauged  Supergravity}
\author{D.H.~Correa\thanks{CONICET}, A.D.~Medina,
E.F.~Moreno\thanks{Associated with CONICET},
\\and \\
F.A.~Schaposnik\thanks{Associated with CICPBA}\\
{\normalsize\it Departamento de F\'\i sica, Universidad Nacional
de La Plata}\\
{\normalsize\it C.C. 67, 1900 La Plata, Argentina}
}
\date{\today}
%\maketitle

%===================================================================
\begin{abstract}
We study monopole and dyon solutions to the equations of motion
of the bosonic sector of ${\cal N} = 4$ gauged supergravity
in four dimensional space-time. A static, spherically symmetric ansatz for
the metric, gauge fields, dilaton and axion leads to soliton
solutions which, in the electrically charged case, have compact
spatial sections. Both analytical and numerical results for the
solutions are presented.
\end{abstract}


\maketitle

\section{Introduction}
%\section{\ Introduction\-}
The interest in gravitating monopole and other soliton solutions
to the equations of motion of non-Abelian gauge theories coupled
to gravity has recently increased  in view of the role that these
solutions may play as backgrounds of gauged supergravity models in
the context of the AdS/CFT correspondence.

Most of the known solutions are available only numerically. An
exception is the self-gravitating monopole solution constructed by
Chamseddine and Volkov (ChV) by solving the first order
Bogomol'nyi (BPS) equations for an ${\cal N}=4$ gauged
supergravity model in four dimensional space
\cite{ChV1}-\cite{ChV2}. Different BPS solutions of the same model
were also found in \cite{Radu1}. The ChV solution corresponds to a
regular magnetic monopole and a geometry which is not
asymptotically flat, the dilaton potential providing a position
dependent negative cosmological term $\Lambda[\phi]<0$. This
solution is purely magnetic and the axion field   is set to zero.
A one-parameter family of non-BPS monopole-like solutions to the
equations of motion of this model was numerically constructed in
\cite{V3}. In some finite range of the parameter $w^{(2)}$, which
is related to the gauge field behavior  at the origin, solutions
are globally regular. Remarkably, when the parameter takes values
in one of the regions outside this range, the corresponding
solutions have compact spatial sections, {\it i.e.} the metric
exhibits a singularity of the  ``bag of gold'' type.

Among other related  self-gravitating Yang-Mills  solutions,
particularly interesting are those found in
\cite{hosr}-\cite{Radu3} for Einstein-Yang-Mills theory (EYM) in
asymptotically anti-de Sitter ($\Lambda <0$) space. In this case,
monopole and dyon solutions are shown to exist (while for $\Lambda
\geq 0$ electrically charged solutions are forbidden
\cite{hosr}-\cite{gal}). Interestingly enough, these dyon
solutions, which have been shown to be stable in some cases,
exhibit non-integer magnetic charge, which shows their
non-topological character. ~

It is the purpose of this work to construct dyon solutions to the
equations of motion in the bosonic sector of ${\cal N} = 4$ gauged
supergravity in four dimensional space (the Freedman-Schwarz model
\cite{FS}). We investigate static, spherically symmetric field
configurations which, in view of the fact they should carry
electric charge, necessarily include a non-trivial axion field
(which for simplicity was set to zero in the purely magnetic
solutions of \cite{ChV1}-\cite{V3}). As we shall see, electrically
charged BPS solutions which are  regular at the origin do not
exist. Then, one has  to study the second order equations of
motion which after the static, spherically symmetric ansatz,
reduce to a system of six coupled non-linear  radial equations
that has to be solved numerically. Again, one finds a family of
solutions now labeled by three parameters: the one already present
for purely magnetic solutions and two new ones related to the
axion and electric field. The properties of these  electrically
charged solutions radically differ from those of the BPS monopole
solution. First, the solutions exhibit a bag of gold behavior
(singular geometry) in the whole range of parameters. After the
study of geodesics for such configurations, one  confirms that the
geometry has a real singularity at some finite value $\rho_*$ of
the ``radial'' coordinate $\rho$. Moreover, in contrast with the
BPS monopole solution  and related to the singular geometry, the
charges have a distinctive behavior and in particular the magnetic
charge is finite but non-integer. Now, in the context of the
AdS/CFT correspondence one should investigate whether the four
dimensional singularity is acceptable or not, following either the
criterion proposed in \cite{Gubser} or uplifting the solution to
10 dimensions and making an analysis similar to that in \cite{MN}.
This requires a thoughtful study on how  the peculiar properties
of dyon solutions manifest in $d=10$, an issue that is left for
future investigations.

The paper is organized as follows. In section 2 we introduce the
model and derive the equations of motion with the spherically
symmetric ansatz. We also define the conserved electric and
magnetic charges. Purely magnetic and dyon solutions are discussed
in sections 3 and 4 respectively. The analysis of the asymptotic
behavior  of the solutions leads to a precise characterization of
bag of gold singularities and how they affect the resulting values
of magnetic and electric charges. Finally, in section 5 we give a
summary of our results.

\section{The model}
Let us start by writing  bosonic part of the $d=4$
Freedman-Schwarz \cite{FS} supergravity model with Lagrangian
\begin{eqnarray}
L \!\!&=& \!\!
\frac {R}{4} -\frac{1}{2}
\partial_\mu\phi\partial^\mu\phi -
\frac{1}{2} \exp(-4\phi) \partial_\mu {\bf a}\partial^\mu {\bf a} -
\frac{\exp(2\phi)}{4}\!\! \sum_{(\sigma)=1,2}
\frac{1}{g^2_{(\sigma)}}  F^{(\sigma) a\,\mu\nu}
F^{(\sigma) a }_{\mu \nu} \nonumber\\
&& - \frac{1}{2} {\bf a} \sum_{{(\sigma)} = 1,2}
\frac{1}{g^2_{(\sigma)}}  {^*}\!F^{(\sigma) a\,\mu\nu}
F^{(\sigma) a }_{\mu \nu} +
\frac{1}{8} \left( g_{(1)}^2 + g_{(2)}^2
\right) \exp(-2\phi)
\label{uno}
\end{eqnarray}
We have chosen the signature of the space-time metric as $(-1,1,1,1)$ and
the convention for the Riemann and Ricci tensors are
$R^\alpha_{\,\beta\mu\nu} = \partial_\mu \Gamma^\alpha_{\beta\nu} - \ldots $
and $R_{\mu \nu} = R^\alpha_{\, \mu \alpha \nu}$. The gauge group is the
direct product $SU(2) \times SU(2)$ and we call $A_\mu^{(\sigma) a}$ the
corresponding gauge connections ($(\sigma) =1,2$). The field-strength is
defined
as
\be
F_{\mu\nu}^{(\sigma) a} = \partial_\mu A_\nu^{(\sigma) a} - \partial_\nu
A_\mu^{(\sigma) a} + \varepsilon_{abc} A_\mu^{(\sigma) b}A_\nu^{(\sigma) c}
\label{2}
\ee
The dual tensor is defined as
\be
{^*}\!F_{\mu\nu}^{(\sigma) a} = \frac{1}{2} \sqrt{|g|}
\varepsilon_{\mu \nu \alpha \beta} F^{(\sigma) a \,\alpha \beta}
\label{3}
\ee
As usual, we denote  $\phi$ the dilaton field and ${\bf a}$ the axion.

We  now consider  Lagrangian (\ref{uno}) when the model is
truncated so that $A_\mu^{(2) a} = 0$ (and $g_{(2)} = 0$) and
write the corresponding equations of motion (omitting the
$(\sigma) = 1$ index)
\begin{eqnarray}
&&{\nabla}_\mu \nabla^\mu \phi = \frac{1}{2g^2} \exp(2\phi) F^a_{\mu\nu}
F^{a\,\mu\nu} -2 \exp(-4\phi) \partial_\mu {\bf a} \partial^\mu{\bf a}
+ \frac{1}{4} g^2 \exp(-2\phi) \nonumber\\
&&\nabla_\mu \left(\exp(2\phi) F^{a\,\mu \nu}
\right) + \exp(2\phi) \varepsilon_{abc} A_\mu^b F^{c \mu \nu}= 2
{^*}\!F^{a\,\nu\mu} \partial_\mu{\bf a}\nonumber\\
&& R_{\mu\nu} = 2 \partial_\mu \phi \partial_\nu\phi +
2 \exp(-4\phi)  \partial_\mu{\bf a} \partial_\nu{\bf a}
+ \frac{2}{g^2} \exp(2\phi)
\left(\vphantom{\frac{1}{4}}F^{a}_{\,\mu\alpha} F^{a \; \alpha}_{~\,\nu}
-\right. \nonumber\\
& & \hphantom{R_{\mu\nu} =}\left.
 \frac{1}{4}g_{\mu\nu} F^{a}_{\,\alpha\beta } F^{a \, \alpha\beta }
\right) - \frac{g^2}{4}g_{\mu \nu} \exp(-2\phi)   \nonumber\\
&&{\nabla}_\mu \nabla^\mu\left(\exp(-4\phi) {\bf a}
\right)  = -  \frac{1}{2g^2} {^*}\!F^{a}_{\,\alpha\beta } F^{a \, \alpha\beta }
\label{cuatro}
\end{eqnarray}

As mentioned in the introduction, Chamseddine and Volkov  found
exact static solutions for the bosonic sector of the truncated
Freedman-Schwarz model
 for purely magnetic gauge fields, $A_0^a$ and ${\bf a} = 0$
\cite{ChV1}-\cite{ChV2}.  In fact, first order Bogomol'nyi
equations \cite{Bogo} were obtained by analyzing the equations for
Killing spinors  and then an exact globally regular solution (also
solving the second order equations (\ref{cuatro})) was found.
Contrary to the expectation that a neutral solution was to be
obtained since the model has no Higgs field, the behavior of the
gauge field solution corresponds to a regular magnetic monopole.
The non-trivial dilaton field provides a potential supporting such
a solution, implying that the resulting geometry is not
asymptotically flat. Although the absence of the Higgs scalar
prevents the definition of a gauge invariant magnetic field, one
can define a magnetic flux which for the Chamseddine-Volkov exact
solution takes the unit value.

Finding dyon solutions  ({\it i.e.} both magnetically and
electrically charged) to eqs. (\ref{cuatro}) is considerably more
complicated since the existence of non-trivial electric field
strength components prevents to set the axion field to zero. Even
the search of more tractable first order Bogomol'nyi equations
becomes more involved since the compatibility condition for the
Killing spinor equations, now including an axion field is highly
non-trivial.

Let us start by proposing a static  spherically symmetric   ansatz
for the space-time metric and the dilaton, axion and gauge fields,
\begin{eqnarray}
&&\!\!\!\!\!\!\!\!ds^2  =   -N(r)\sigma^2(r) dt^2 + \frac{1}{N(r)} dr^2 + r^2 \left(
d\theta^2 + \sin^2 \theta d\varphi^2
\right)\label{ans0}\\
&&\!\!\!\!\!\!\!\!A =A_\mu^a t^a dx^\mu =
u(r) t^3 dt + w(r) (-t^2 d\theta + t^1 \sin \theta d\varphi) + t^3 \cos \theta
d\varphi
 \label{ans1}
\end{eqnarray}
\be
\phi =  \phi(r) \, ,    \;\;\;\;\;\;\;   {\bf a} = a(r)
\label{ans2}
\ee
where  we have set $g=1$ without loss of generality. Let us note
that the static spherically symmetric ansatz (\ref{ans1}) for the
gauge field  can be obtained from the general one introduced in
\cite{witt}-\cite{man} by a singular gauge transformation (see
\cite{hos} for details) so that $w(r)$  can be associated with
that introduced in the case of the  't Hooft-Polyakov monopole
\cite{tH}-\cite{Pol} through the ansatz $A_i^a = (1 -
w)\varepsilon_{aij} x^j/r^2$ and $u$ with the scalar potential for
the Julia-Zee dyon \cite{JZ}, $ A_0^a = u(r) x^a/r$.

Inserting ansatz (\ref{ans0})-(\ref{ans2}) into (\ref{cuatro}) one can write
six   independent radial equations of motion in the form
\begin{eqnarray}
&&   r^2 \exp(-4 \phi) \sigma^2 N^2 a'^2
+ r^2 \sigma^2  N^2 {\phi'}^2 - r \sigma N^2 \sigma'
+ 2 \exp(2\phi) u^2 w^2  + \nonumber\\
 &&  2 \exp(2\phi) \sigma^2 N^2 {w'}^2 = 0
 \label{primera}
 \\
  &&  4 r^3 \sigma N \sigma' + 4r^2 \sigma^2
(-1 + N + rN') +
4 \exp(2\phi) \sigma^2 (w^2 - 1)^2  + \nonumber  \\  &&
4 \exp(2\phi) r^4 {u'}^2 -r^4 \sigma^2 \exp(-2\phi)  = 0
\label{segunda}\\
& & \sigma^2 N( 2 N w' \phi' +  w' N' +   N w'')
+u^2 w  + \sigma \sigma' N^2  w'  +
\sigma^2 N w \frac{1}{r^2} (1-w^2)  +\nonumber
\\
& & 2 \sigma  N u w a' \exp(-2\phi)
  = 0 \label{tercera} \\
  && r^2 \phi''  N \sigma +
\phi' (r^2 N' \sigma  + 2r  N \sigma + r^2 N \sigma') - \frac{r^2}{4} \sigma
\exp(-2\phi)  + \nonumber
\\
&& 2 r^2 \exp(-4\phi)\sigma N {a'}^2  +2  \sigma^{-1}N^{-1} u^2
w^2 \exp(2 \phi) + r^2 \sigma^{-1} {u'}^2 \exp(2\phi) -
\nonumber\\ && \frac{1}{r^2} \sigma (w^2 - 1)^2 \exp(2\phi)  - 2
\sigma N{ w'}^2 \exp(2 \phi) = 0
\label{cuarta}\\
&& 2 \sigma^2 N (w^2 - 1)  a' \exp(-2\phi) + r^2 N (\sigma' u' +
\sigma u'') + 2 \sigma u w^2 - \nonumber
\\
&& 2 r N \sigma u' (1 + r \phi') = 0
\label{quinta}\\
& & r^2 N\sigma a'' +  r \sigma N  \left( a'(2 - 8 r \phi')
+ 4 a (-2 \phi' + 4 r {\phi'}^2 - r \phi'')
\right) + \nonumber\\
&& r^2 (\sigma N)' (a' - 4 a \phi')   + 2 \exp(4\phi) \left(
(w^2 - 1) u' + 2 u w w'
\right) = 0
\label{ultima}
\end{eqnarray}

\subsection*{Defining magnetic and electric charges}
As already noted in \cite{ChV1}-\cite{ChV2}, defining a magnetic
charge in the Freedman-Schwarz model is problematic since there is
no Higgs field breaking the gauge symmetry thus providing a
natural isospin direction to project the $SU(2)$ field strength on
the direction of the residual Abelian symmetry (as one does for
the original flat space  't Hooft-Polyakov (tHP) monopole
configuration). Note however that ansatz (\ref{ans1}) is nothing
but the gauge-transformed  of the original tHP ansatz (we call
$\bar A$ the gauge field ansatz in its original tHP form)
entangling space and isospace indices,
\ba
A_i &\to& \bar A_i = S^{-1}A_i S + \frac{i}{e} S^{-1} \partial_i S =
 \frac{i}{e} (1-w)
 [\Omega, \partial_i\Omega]\nonumber\\
 A_0 &=& S^{-1}A_0 S = u(r) \Omega\nonumber\\
\Omega &=&  {t^a}  \frac{x^a}{r}
\label{nonumber}
\ea
here $S$ is given by
\be
S= \left( \matrix{
~~~\exp\left(i (\phi +\eta)/2\right) \cos(\theta/2)  &
\exp\left(-i (\phi -\eta)/2\right) \sin(\theta/2) \cr
-\exp\left(i (\phi -\eta)/2\right) \sin(\theta/2) &
~\,~\exp\left(-i (\phi +\eta)/2\right) \cos(\theta/2) \cr}
\right)
\label{newmetric}
\ee
where $\eta$ is some function of $(r,t)$ to be appropriately
chosen \cite{hos}.

Then, as for the pure Yang-Mills pioneering monopole ansatz of Wu
and Yang \cite{YW}, one can define a projected field strength in
the form
\be
 {\cal F}_{\mu\nu}  = {\rm Tr} \left(\bar F_{\mu\nu}  \Omega \right)
\label{u1}
\ee
which, written in terms of fields in the original gauge becomes
\be {\cal F}_{\mu\nu}  = {\rm Tr} \left( F_{\mu\nu} S \Omega
S^{-1}\right) \label{cal} \ee
leading to a   magnetic field ${\cal B}^i = (1/2\sqrt{-g})
\varepsilon^{ijk}{\cal F}_{ij} $ which takes the form
\be
{\cal B}^r = \frac{1}{\sqrt{-g}} (1 - w^2) \label{1-w}
\label{Br}
\ee
and an electric field ${\cal E}^i = {\cal F}^{0i}$ of the form
\be
{\cal E}^r = \frac{u'(r)}{\sigma^2}
\ee
Now, we define the electric and magnetic charges in the form
\begin{eqnarray}
 Q_M &=& \frac{1}{4\pi} \int dS_k  \sqrt{-g} {\cal B}^k\nonumber\\
Q_E &=& \frac{1}{4\pi} \int dS_k  \sqrt{-g} {\cal E}^k
\label{cargas}
\end{eqnarray}
so that for ansatz (\ref{ans0})-(\ref{ans1}) one has

\begin{eqnarray}
Q_M &=&   1 - w(\infty)^2
\label{cargam}\\
Q_E &=& \left.\frac{u' r^2}{\sigma} \right|_\infty \label{cargae}
\end{eqnarray}

As stated in the introduction, exact solutions of the first order
Bogomol'nyi equations also solving the second order equations. of
motion of the truncated Freedman-Schwarz model are known
\cite{ChV1}-\cite{ChV2}. They correspond to   mono\-pole-like
gauge field configurations (with magnetic charge 1 and zero
electric charge) and a globally hyperbolic regular  geometry.
Apart from these exact solutions, other magnetically charged
non-BPS solutions were found in \cite{V3}. Let us then  analyze,
for completeness, this well-studied purely magnetic case and then
discuss dyon solutions.

\section{Non-BPS monopole solutions}

We then start from eqs.~(\ref{primera})-(\ref{ultima})  with
$u=a=0$, thus reducing the system to one with 4 coupled nonlinear
equations. Now, as pointed in \cite{ChV1}-\cite{ChV2}, the bosonic
action admits a global symmetry which in our notation corresponds
to
\be
\phi \to \phi - \epsilon \, , \;\;\;\;  N \to\exp(2\epsilon)  N
\label{simetria}
\ee
Following the Noether theorem, there is a conserved current which
leads to the following relation between  $N,\ \sigma$ and $\phi$,
\be
N \sigma^2 = \exp\left(2(\phi - \phi_0) \right)
\label{relacion} \ee
where $\phi_0$ is an integration constant. Using this relation,
one can reduce the coupled  system of 4 equations of motion to 3
independent equations for 3 unknown functions,
\begin{eqnarray}
r &\to& \rho \nonumber\\
(w,\phi,N,\sigma ) &\to& (w,\phi,R)
\label{sinnombre}
\end{eqnarray}
where we have chosen  $R$ as the third unknown function,
 implicitly defined while
introducing a new variable $\rho$ replacing the original radial variable $r$
through
\begin{equation}
r^2 =  \exp\left(2(\phi-\phi_0)\right) R^2(\rho)
\label{rro}
\end{equation}
This change is inspired in that yielding to the exact ChV solution
\cite{ChV1}-\cite{ChV2}. It yields to a metric of the form
\be ds^2 = \exp(2\phi - 2 \phi_0) (-dt^2 + d\rho^2 + R^2(\rho)
d\Omega^2) \label{nuevamet0} \ee

After some algebra, the resulting system can be written as
\begin{eqnarray}
& & w'' + 2\phi' w' - \frac{1}{R^2} w \left(w^2 - 1 \right) = 0\nonumber\\
& & \phi'' + 4 {\phi'}^2 + \frac{1}{R^2} \left({R'}^2 -2 {w'}^2 -1\right)
+ 6 \frac{R'}{R} \phi' - 1 = 0 \nonumber\\
& & R'' + \frac{1}{R} \left(3{w'}^2 - {R'}^2 + 1 \right) - 4R {\phi '}^2
+ R - 6 \phi'R' = 0
\label{tres}
\end{eqnarray}
where $w' = dw/d\rho$.

These equations together with the regularity condition at the origin
imply
\begin{eqnarray}
w &=& 1 - w^{(2)} \rho^2 + {\rm O}(\rho^4) \nonumber\\
\phi &=& \phi^{(0)}  + \phi^{(2)} \rho^2 + {\rm O}(\rho^4) \nonumber\\
%\sigma &=&  1  + \sigma^{(2)} r^2 + {\rm O}(r^4) \nonumber\\
%N &=& 1  + N^{(2)} \rho^2 + {\rm O}(\rho^4) \nonumber\\
R &=& \rho - R^{(3)} \rho^3 + {\rm O}(\rho^5)
\label{origen}
\end{eqnarray}
Here $w^{(2)}$ and $\phi^{(0)}$ are free parameters. Note that the
equations of motion  are invariant under the transformation
\begin{eqnarray}
& & \rho \to \lambda \rho  \; , \;\;\; \;\;\; \;\;\;R \to \lambda R\nonumber\\
& & w \to w \; , \;\;\; \;\;\; \;\;\;\;
\phi \to \phi + \log \lambda
\label{shift}
\end{eqnarray}
and hence the value of $\phi^{(0)}$ can be arbitrarily fixed.  In
order to compare with the BPS solution in \cite{ChV1}-\cite{ChV2},
we choose $\phi^{(0)}= -\log 2/2$. Then eqs. (\ref{tres}) imply
the following relation between coefficients $\phi^{(2)}$,
$R^{(3)}$ and $w^{(2)}$,
\begin{eqnarray}
\phi^{(2)} &=&  w^{(2)\,2} + \frac{1}{12} \nonumber\\
R^{(3)} &=& w^{(2)2} + \frac{1}{36}
\label{rela2}
\end{eqnarray}
We are then left with just one shooting parameter $w^{(2)}$.

Concerning spatial infinity, we seek for solutions having the
following asymptotic behavior
\begin{eqnarray}
w  &\sim &    w_1^{(\infty)} \frac{1}{  \rho^{1/2}} + w_2^{(\infty)}
\frac{1}{  \rho^{3/2}} + \ldots + C\rho\exp(-\rho) + \ldots
\nonumber\\
\phi & \sim & \phi_1^{(\infty)} + \frac{\rho}{2} - \frac{1}{4} \log \rho
+ \phi_2^{(\infty)}\frac{1}{\rho^2} + \ldots  + D\sqrt x\exp(-\rho) + \ldots
\nonumber\\
R &\sim& \sqrt{2\rho} - R_1^{(\infty)} \frac{1}{  \rho^{3/2}}
\label{infinity}
+ \ldots  + F \rho\exp(-\rho) + \ldots
\end{eqnarray}
With this behavior the magnetic charge, as given by
(\ref{cargam}), is $Q_M = 1$.

Using these conditions, solutions can be found numerically,
varying the shooting parameter $w^{(2)}$ in a given range of
values and analyzing the resulting asymptotic behavior. We have
found a whole range $R_m$  of values for $w^{(2)}$ for which
globally regular  solutions exist, which coincides with  that
found in \cite{V3}, $R_m = (w^{(2)}_a=0,w^{(2)}_b=1/2)$. For the
particular value $w^{(2)} =1/6$, the solution coincides with the
ChV exact one. For $w^{(2)}_b>1/2$, as first shown in \cite{V3},
one finds that $R$ develops a second zero (apart from that at
$\rho = 0$) at some finite value $\rho_*$ where the geometry is
singular. These kind of solutions are similar to those called bags
of gold \cite{wheeler} and are also related to the black hole
solutions found in \cite{V3}. We shall describe in more detail
this behavior below.

Typical  solutions are displayed in Figs.~1-2. The shape of
solutions for  $\phi,w, R$ as  functions of $\rho$ in the whole
range $R_m$ is similar to that of the exact ChV solution.
Concerning the gauge field, one finds that   certain solutions,
those corresponding to $w^{(2)}>1/6$ develop nodes (as it also
happens for the Einstein-Yang-Mills monopole solutions in
asymptotically AdS space, found in \cite{hos}). The magnetic
charge, in view of the asymptotic behavior of $w(\rho)$, has value
one for $w^{(2)}<1/2$ but becomes non-integer for $w^{(2)}>1/2$ so
that the configuration corresponds in this case to a kind of
non-topological soliton  (we shall discuss more in detail this
issue in the next section). Finally, note  that, except the
solution corresponding to $w^{(0)} = 1/6$,   all other solutions
to the second order equations of motion do not satisfy first order
Bogomol'nyi equations.

\begin{figure}%[ht]
\centerline{
\psfig{figure=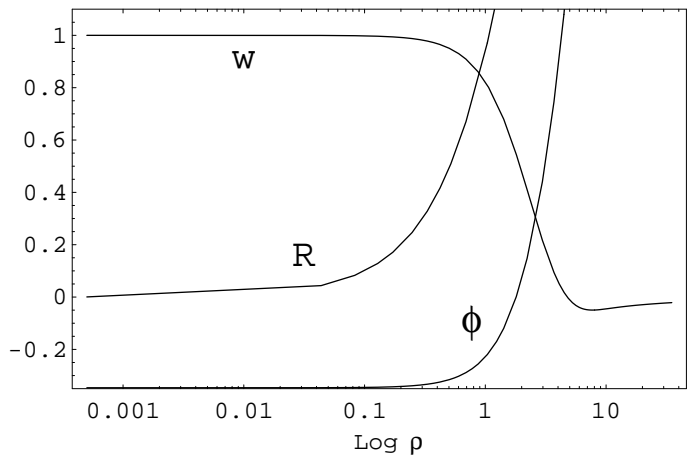,height=6cm,angle=0}}
\smallskip
\caption{Globally regular purely magnetic (non-BPS) solution for
$w^{(2)} = 0.2$ \label{fig1} }
\end{figure}

\begin{figure}%[ht]
\centerline{
\psfig{figure=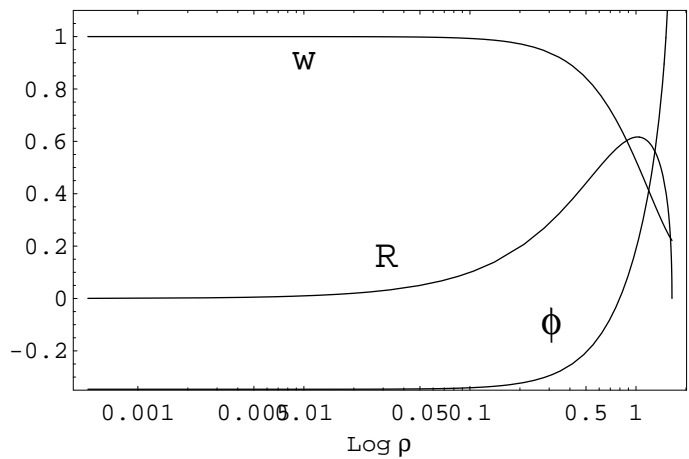,height=6cm,angle=0}}
\smallskip
\caption{ ``bag of gold'' purely magnetic
solution for $w^{(2)} = 0.7$ \label{fig2} }
\end{figure}

Let us now study more in detail the behavior of the bag of gold
solutions, which in the present purely magnetic  configurations
correspond to $w^{(2)} >1/2$. We start by defining a radius $R_c$
as the function multiplying the solid angle $d\Omega^2$ of the
spatial geometry (see eq. (\ref{nuevamet0})),
\begin{equation}
R_c^2(\rho) = \exp(2\phi) R^2(\rho)
\label{radio}
\end{equation}
For bag of gold solutions  $R_c$ should vanish at some $\rho =
\rho_*$,  $R_c(\rho_*) =0$. To see this, one has to determine the
behavior of $R(\rho)$ and $ \exp(2\phi)$ when $\rho \sim \rho_*$.
This can be done analytically by proposing an asymptotic  behavior
in powers of $x = \rho_*- \rho$, finding out that
\begin{eqnarray}
w \sim {\rm constant}  \; , \;\;\;\; \exp(\phi) \sim C x^{-1/4}
\
\; , \;\;\;\; R \sim
D x^{1/2}
\label{constant}
\end{eqnarray}
so that one has indeed
\be
R_c^{mon\, 2}(\rho) \sim  D^2C^2 (\rho_* - \rho)^{1/2}
\label{levoyaponer}
\ee
which corresponds to the behavior of a bag of gold solution.

\section{Dyon solutions}

Let us now extend our analysis to the electrically charged case.
Being $a \ne 0$ and $u \ne 0$, symmetry (\ref{simetria}) is lost
and hence no relation of the kind (\ref{relacion}) can be
established. We are then left with six equations of motion for six
unknown functions.

One could try to find BPS solutions by analyzing the supersymmetry
variations of fermionic fields when $a$ and $u$ fields are
included. The conditions resulting from the vanishing of these
variations  are highly more involved than those arising in the
purely magnetic case.  However, by assuming a regular behavior at
the origin, we found that the only non-trivial solution to the
first order system corresponds to the exact purely magnetic
solution discovered in \cite{ChV1}-\cite{ChV2}. We are then forced
to study the second order equations of motion to determine whether
non-BPS dyon solutions exist.

In analogy with the purely magnetic case, we shall write the
system  in terms of a new variable $\rho$ and   unknown functions
in the form
\begin{eqnarray}
r &\to& \rho \nonumber\\
(w,u,\phi,a,N,\sigma) &\to& (w,u,\phi,a,R,V)
\label{relacion2}
\end{eqnarray}
with
\begin{eqnarray}
&& \exp(2V)  =  N\sigma^2 \nonumber\\
 &&\exp(2V) R^2(\rho) = r^2
 \label{sinnombre2}
\end{eqnarray}
After the change (\ref{relacion2}), the metric (\ref{ans0}) becomes
\be
ds^2 = \exp(2V) (-dt^2 + d\rho^2 + R^2(\rho) d\Omega^2)
\label{nuevamet}
\ee
Note finally that in the purely magnetic case, due to the global
symmetry (\ref{simetria}) one can identify $V$ with $\phi$
according to the relation   $V = \phi + \log2/2$ and then
(\ref{nuevamet}) becomes (\ref{nuevamet0}).

The system of equations can then be brought to the form
\begin{eqnarray}
& & w'' + 2\phi' w' - \frac{1}{R^2} w \left(w^2 - 1 \right) + u^2 w
+ 2 \exp(-2\phi) u w a'= 0\nonumber\\
& & \phi'' + 2 \phi' (V' +  \frac{R'}{R}) + \exp (2\phi - 2V)
\left({ u'}^2  - 2 \frac{{w'}^2 - u^2 w^2}{R^2}
- \frac{(1 - w^2)^2}{R^4}  \right)  \nonumber\\
&&
-\frac{1}{4} \exp (-2\phi + 2V)  + 2 \exp(-4\phi){a'}^2 = 0 \nonumber\\
&& V'' + 6 \frac{R'}{R} V' + 5 {V'}^2 -4  \exp (2\phi - 2V) \left(
\frac{{w'}^2}{R^2} + \frac{w^2 u^2}{R^2}
\right) - \frac{1}{R^2} -\nonumber \\
&& {\phi'}^2 -\exp(-4\phi) {a'}^2 + \frac{{R'}^2}{R^2} -
\frac{1}{2} \exp\left(-2\phi + 2 V\right) = 0 \nonumber\\
&& R'' - 6 V'R'
-\frac{{R'}^2}{R} +\frac{1}{2}\exp(-2\phi +2V)R
+ 2R ({\phi'}^2 - 3{V'}^2) + \nonumber \\
&& 6 \exp(2\phi - 2V)
\left(\frac{{w'}^2 + u^2 w^2}{R}
\right) + \frac{1}{R} + 2 \exp(-4\phi) R {a'}^2 = 0
\nonumber\\
&& -u'' - 2u'\left( {\phi'}  + \frac{R'}{R}\right)+ \frac{2uw^2}{R^2}
    + \frac{2}{R^2}\exp(-2\phi) a' (w^2 - 1)
= 0\nonumber\\
&& a'' + 2a'\left( V' + \frac{R'}{R} - 4 \phi'
\right)  - 8a\phi' \left( \frac{R'}{R} + V' - 2{ \phi'}
\right) + \nonumber\\
&&  \frac{2}{R^2} \exp (4\phi - 2V)\left(
(w^2 - 1)u' + 2 w w' u\right) - 4 a \phi'' = 0
\label{seis}
\end{eqnarray}

It is useful to notice that, apart from the
generalization of invariance (\ref{shift}) already present in the
purely magnetic case,
\begin{eqnarray}
&&\rho \to  \lambda \rho \nonumber\\
&& \phi \to \phi + \log \lambda \, , \;\;\;\;\;
 u \to \lambda^{-1} u  \, , \;\;\; \;\;
w \to w\nonumber\\
&& a \to  \lambda^2 a  \, \;\;\; \;\; V \to V \, , \;\;\;\;\; R
\to  \lambda R \label{seisuno}
\end{eqnarray}
system
(\ref{seis}) exhibits now a second invariance
\begin{eqnarray}
&&\rho \to  \mu \rho \nonumber\\
&& \phi \to \phi   \, , \;\;\;\;\;
 u \to \mu^{-1} u  \, , \;\;\; \;\;
w \to w\nonumber\\
&& a \to   a  \, , \;\;\; \;\; V \to V - \log \mu  \, , \;\;\;\;\; R
\to  \mu R \label{seisdos}
\end{eqnarray}
and this reduces the number of shooting parameters to fix when
seeking for a numerical solution.

%%%%%%%%%%%%%%%%%%%%%%%%%%%%%%%%%%%%%%%%%%%%%%

 ~

%%%%%%%%%%%%%%%%%%%%%%%%%%%%%%%%%%%%%%%%%%%%%%

For the case of dyon  solutions, we have to supplement
(\ref{origen}) with conditions at the origin for $u$, $a$ and $V$,

\begin{eqnarray}
u &=& u^{(0)} \rho + u^{(3)} \rho^3 + {\rm O}(\rho^5) \nonumber\\
a &=& a^{(0)}  + a^{(2)} \rho^2 + {\rm O}(\rho^4) \nonumber\\
V &=& V^{(2)} \rho^2 + {\rm O}(\rho^4)
\label{origin2}
\end{eqnarray}
Note that, in view of symmetry (\ref{seisdos}) one can fix $V(0) = 0$.
Moreover, eqs. (\ref{seis}) now allow to write $a^{(2)}, V^{(2)},
u^{(3)}, f^{(2)}$ and $R^{(3)}$ in terms of $w^{(2)}, u^{(0)}$ and $a^{(0)}$,
\begin{eqnarray}
a^{(2)} &=& \frac{1}{6}\left( 2a^{(0)} - 6a^{(0)}u^{(0)\,2} -
3u^{(0)} w^{(2)} + 24 a^{(0)}
w^{(2)\,2}
\right) \nonumber\\
V^{(2)} &=& \frac{1}{12}\left(1 + 3 u^{(0)\,2} + 12 w^{(2)\,2}
\right) \nonumber\\
f^{(2)} &=& \frac{1}{12}\left(1 - 3 u^{(0)\,2} + 12 w^{(2)\,2}
\right)\nonumber\\
R^{(3)} &=& \frac{1}{36} \left(1 + 9 u^{(0)\,2} + 36 w^{(2)\,2}
\right) \nonumber\\
u^{(3)} &=& \frac{1}{90} \left( -  u^{(0)} + 27  u^{(0)\,3} + 48
a^{(0)}w^{(2)} + 36 u^{(0)} w^{(2)} - 144 a^{(0)} u^{(0)\,2}w^{(2)} -
\right.
\nonumber\\
&& \left.
36 u^{(0)} w^{(2)\,2} + 576 a^{(0)} w^{(2)\,3}\right)
\label{relacd}
\end{eqnarray}
Using these conditions one can numerically integrate eqs.
(\ref{seis}), with $w^{(2)}, u^{(0)}$ and $a^{(0)}$ as shooting
parameters. As we shall describe in detail below, the space of
dyon solutions radically changes with respect to the pure monopole
case discussed above. Indeed, all dyon solutions   are ``bag of
gold'' type: function $R(\rho)$ always has two zeros -one at $\rho
= 0$, the second one at some finite value $\rho_*(w^{(2)},
u^{(0)},a^{(0)})$. One can see numerically that this happens for
all positive $ w^{(2)}$ values, (negative ones lead to an
unbounded magnetic field at $\rho = \rho_*$).

To see this in more detail, let us start by considering the region
$0< w^{(2)} <1/2$ (where regular pure monopole solutions were
found). For small values of $ u^{(0)}$  and $a^{(0)}$ ($u^{(0)}
\sim a^{(0)}\simeq 0.1$)   all $(w,R,V,\phi)$ solutions are, at
small $\rho$,  very similar to the monopole ones but, as one
increases $\rho$, all except $w$ start deviating from such a
behavior. Indeed, the behavior of $w$ for the dyon solution is
much like that of the monopole solution in the whole range
$0\leq\rho\leq\rho_*$. In contrast, although  $R$,  starts growing
as in the monopole solution, it reaches in this case a maximum
value and then decreases monotonically, until it vanishes at some
fixed value $\rho_*$. As we already discussed in the case of the
monopole solution (for $w^{(2)}
>1/2$), this behavior corresponds to
a bag of gold. We have found that  the electric and axion fields,
$u(\rho)$ and $a(\rho)$, have no nodes  for $w^{(2)} <1/6$ while
in the region $w^{(2)} > 1/6$ they have a growing number of nodes.
Typical behaviors are shown in figures 3 and 4.

%%%%%%%%%%%%%%%%%%%%%%%%%%%%%%%%%%%%%%%%%%%%%%%%%%%%%%%%%%%%%%%
\begin{figure}%[ht]
\centerline{
\psfig{figure=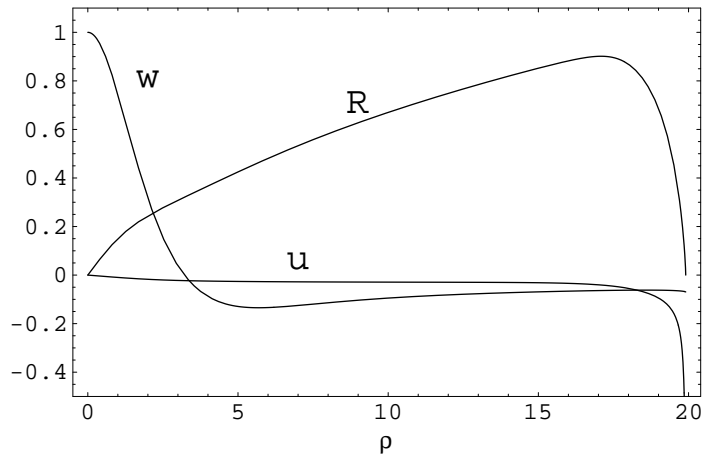,height=6cm,angle=0}}
\smallskip
\caption{Dyon solution for gauge field and $R$ for $ w^{(2)} =
0.3, u^{(0)} = -0.01, a^{(0)} = 0.00414 $ \label{fig3} }
\end{figure}
%%%%%%%%%%%%%%%%%%%%%%%%%%%%%%%%%%%%%%%%%%%%%%%%%%%%%%%%%%%%%%%

%%%%%%%%%%%%%%%%%%%%%%%%%%%%%%%%%%%%%%%%%%%%%%%%%%%%%%%%%%%%%%%
\begin{figure}%[ht]
\centerline{
\psfig{figure=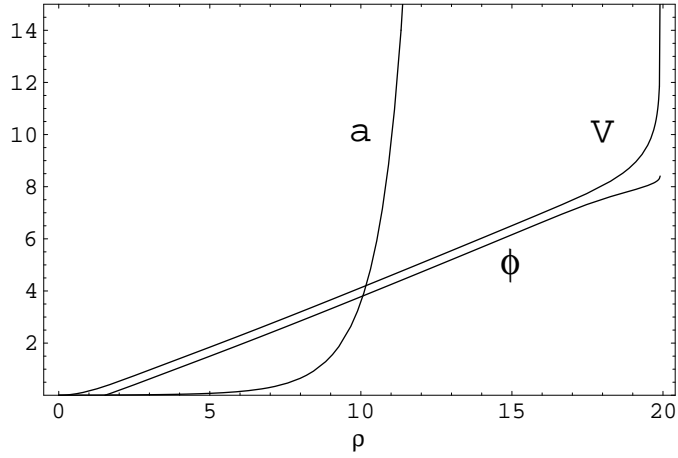,height=6cm,angle=0}}
\smallskip
\caption{Axion, dilaton and $V$ solutions for
 $ w^{(2)} = 0.3, u^{(0)} = -0.01, a^{(0)} = 0.00414 $ \label{fig4} }
\end{figure}
%%%%%%%%%%%%%%%%%%%%%%%%%%%%%%%%%%%%%%%%%%%%%%%%%%%%%%%%%%%%%%%

As one approaches $ w^{(2)} = 1/2$ (the critical monopole solution
value), $w,u$ and $a$ develop more and more nodes (this
oscillatory behavior was already present for $w$ in the monopole
case). For $ w^{(2)} > 1/2$ where also the pure monopole solution
was bag of gold type one finds that, adjusting the shooting
parameters $a^{(0)}$ and $u^{(0)}$  one can make $\rho_*^{dyon} >
\rho_*^{monopole}$. This behavior is shown in figure 5 for
$w^{(2)} = 0.7$. As announced, we see by comparing figures 2 and 5
that $\rho_*$ has grown from $\rho_* = 1.647$ for the purely
magnetic solution to $\rho_* = 3.384$ for the dyon solution.

   ~

%%%%%%%%%%%%%%%%%%%%%%%%%%%%%%%%%%%%%%%%%%%%%%%%%%%%%%%%%%%%%%%
\begin{figure}%[ht]
\centerline{
\psfig{figure=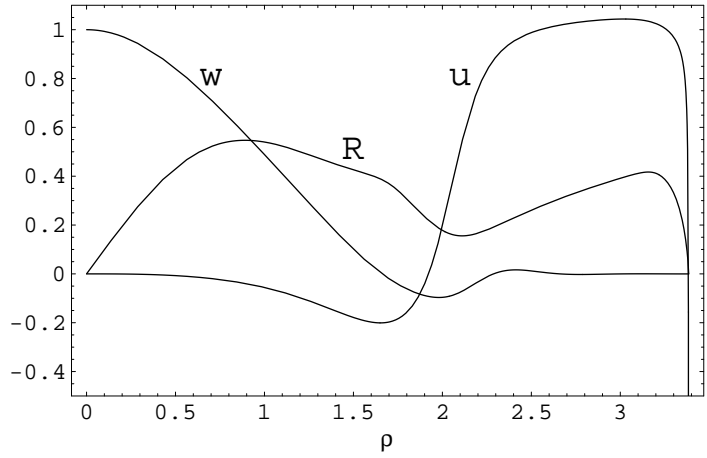,height=6cm,angle=0}}
\smallskip
\caption{Dyon solution for $ w^{(2)} = 0.7, u^{(0)} = -0.01,
a^{(0)} = 0.233 $ \label{fig5} }
\end{figure}
%%%%%%%%%%%%%%%%%%%%%%%%%%%%%%%%%%%%%%%%%%%%%%%%%%%%%%%%%%%%%%%

Let us extend the analysis of purely magnetic bag of gold
solutions presented at the end of section 3 to the case of dyons.
The analytic determination of  asymptotic behaviors   becomes in
this case  more involved. One can however find  bounds which can
be used to study crucial properties of the solutions and can also
be confronted with the results obtained numerically. We start by
assuming  the following asymptotic behavior,

\begin{eqnarray}
w \sim w_0    \; , &&  u \sim B x^\beta
\nonumber \\
\exp(- V) \sim K x^\delta \; , && R \sim A x^\alpha
\label{behave}
\end{eqnarray}
Using these expressions in  the equations  for $V$ and $R$
(\ref{seis}) and requiring the dominant terms to be order $x^{-2}$
(the only consistent possibility), one gets the following
conditions relating exponents $\alpha$ and $\delta$,
\begin{eqnarray}
 \left( \delta - 6 \alpha \delta + 5 \delta^2 + \alpha^2\right)
 x^{-2} - 4 F^2(\rho)
- G^2(\rho) - H^2(\rho)   &=& 0 \nonumber\\
 \left( -\alpha + 6 \alpha \delta - 6 \delta^2 \right)
 x^{-2}+ 6 F^2(\rho)
+ 2 G^2(\rho) + H^2(\rho)   &=& 0
\label{nuevasrel}
\end{eqnarray}
where
\begin{eqnarray}
F^2   &=& \frac{1}{R^2} {\exp(2\phi - 2 V)} ({w'}^2 + w^2 u^2) \nonumber\\
G^2 &=& {\phi '}^2 + \exp(-4\phi) {a'}^2 \nonumber\\
H^2 &=&\frac{1}{2} \exp(-2\phi + 2 V)
\label{reales}
\end{eqnarray}
can at most be of order $x^{-2}$. Consistency of eqs.
(\ref{nuevasrel}), together with reality of $F$, $G$ and $H$,
impose certain conditions  on exponents $\alpha$ and $\delta$. One
of the possibilities that results from (\ref{nuevasrel}) is
\begin{eqnarray}
\alpha \geq \delta \; , \;\;\; \alpha + \delta \leq 1
\label{desi}\\
\alpha \leq 2 \delta \; , \;\;\; \alpha - \delta \leq \frac{1}{2}
\label{desibis}
\end{eqnarray}
from which one finds
\be
0 < \alpha \leq \frac{2}{3}  \; , \;\;\;
\;   \;\;\;  0 < \delta \leq \frac{1}{2}
\label{desitris}
\ee
There are other three possibilities which can be discarded by
comparison with the numerical solutions. Indeed two of them lead
to $\alpha\leq\delta$ while the third one implies $\delta
\leq1/4$. Neither of these conditions is verified by the numerical
solutions. We are then left with  (\ref{desi})-(\ref{desitris}) as
sole possibility.

These bounds are very useful in analyzing some crucial properties
of the solutions. Indeed, from eq. (\ref{nuevamet}) and the
analogous of formula (\ref{radio}) for dyons
\begin{equation}
R_c^{dyon\,2}(\rho) = \exp(2V) R^2(\rho)
\label{radio2}
\end{equation}
one can write
\be R_c^{dyon \, 2} \sim \frac{A^2}{K^2} x^{2(\alpha - \delta)}
\label{tal} \ee so that from   (\ref{desi})  one has
\be
R_c^{dyon}(\rho_*)  = 0
\ee
Then, the three parameter family of solutions corresponds to a
geometry characterized by compact spatial sections (of ``radius''
$\rho_*$). In order to have a better understanding of the
properties of such a geometry, let us evaluate the affine length
${\mathcal{L}}$ associated to a massless particle, moving along a
null radial geodesic, from $\rho = 0$ to $\rho = \rho_*$. To this
end, we start from the null geodesic condition (we call $v$ the
affine parameter)
\be
-\left(\frac{dt}{dv}\right)^2 g_{tt} + \left(\frac{d\rho}{dv}\right)^2
g_{\rho\rho} = 0
\label{ii}\ee
``Energy'' conservation implies
\be
\frac{dt}{dv} = \frac{E}{g_{tt}}
\label{EE}
\ee
so that, from (\ref{ii}) one can write
\be
\frac{d\rho}{dv} = \frac{E}{\sqrt{g_{tt}g_{\rho\rho}}} = \frac{E}{\exp(2V)}
\ee
and then the affine length takes the form
\be
{\cal L} =  \int_0^{\rho_*} dv = \frac{1}{E} \int_0^{\rho_*} \exp(2V) d\rho
\ee
Being $\exp(2V)$ regular in the interval $0 \leq \rho <\rho_*$ the
only possibility of having an infinite affine length (so as to
have a geodesically complete space) should come from the
singularity at $\rho = \rho_*$.  Then, using the asymptotic
expression (\ref{behave}) for $V$, one gets \be {\cal L} =
 {\rm constant} - \lim_{\rho \to {\rho_*}} \frac{C^2}{1 - 2\delta}
(\rho_* - \rho)^{1 -2\delta}
\ee
Now, we have found analytically  that  $0 < \delta \leq 1/2$  so
that the affine length is finite and then $\rho = \rho_*$
corresponds to a real singularity of space-time (incomplete
geodesic). Let us end this discussion by analyzing the scalar
curvature ${\cal R}$ at $\rho = \rho_*$. From the explicit form of
the metric (\ref{nuevamet}), one has
\be
{\cal R} =-2 \exp(-2V)R^{-2}
\left(-1+R'^2 + 6 R R' V' +3 R^2(V' + V'') + 2 R R'' \right)
\label{cur}
\ee
Now, using the asymptotic behaviors (\ref{behave}) one can see
that, to leading order,
\be
{\cal R} \sim \left( \alpha^2 -6\alpha \delta
+ 2 \alpha(\alpha-1) + 3 \delta \right) x^{2\delta-2}
\ee
Using the allowed ranges (\ref{desitris}) for $\alpha$ and
$\delta$ as well as the numerical results one can easily see that
the coefficient of $ x^{2\delta-2}$ does not vanish and hence the
curvature diverges at $\rho = \rho_*$ so that   $\rho_*$
corresponds to a scalar curvature singularity.
Concerning  numerical results one can extract the following values
for $\beta, \delta$ and $\alpha$ (for a typical case corresponding
to $w^{(2)} = 0.3$, $u^{(0)}= -0.01$ and $a^{(0)}=0.00414$),
\begin{eqnarray}
\beta \sim 0.39 \; , \;\;\; \delta \sim 0.34 \; , \;\;\; \alpha \sim 0.65
\label{solotres}
\end{eqnarray}
which as we see, agree with the bounds that one can obtain
analytically. We show in figure 6 the function $R_c(\rho)$ for
this solution.

%%%%%%%%%%%%%%%%%%%%%%%%%%%%%%%%%%%%%%%%%%%%%%%%%%%%%%%%%%%%%%%
\begin{figure}%[ht]
\centerline{
\psfig{figure=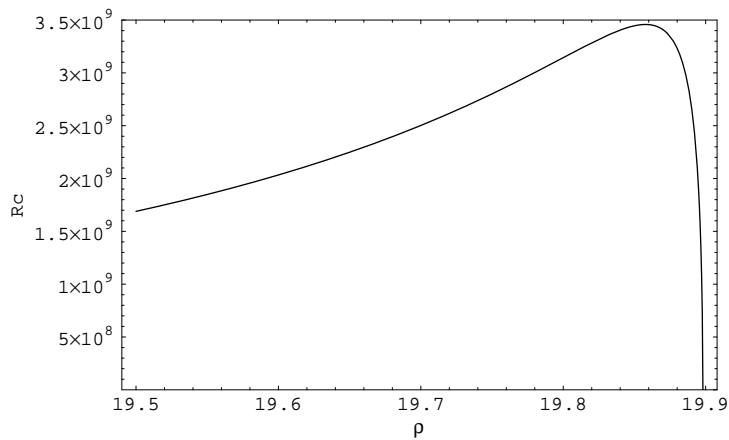,height=6cm,angle=0}}
\smallskip
\caption{
The radius $R_c(\rho)$      for the dyon solution
with $w^{(2)} = 0.3$, $ u^{(0)} = -0.01$, $a^{(0)} = 0.00414 $
\label{fig6antes} }
\end{figure}
%%%%%%%%%%%%%%%%%%%%%%%%%%%%%%%%%%%%%%%%%%%%%%%%%%%%%%%%%%%%%%%

The fact that $R(\rho_*)=0$ corresponds to compact spatial
sections and has important consequences on the magnetic and
electric charges.
%%%%%%%%%%%%%%%%%%%%%%%%%%%%%%%%%%%%%%%%%%%%%%%%%%%%%%%%%%%%%%%%%%%%%%%%%%
Indeed, in order to compute the electric and magnetic charges, one
should use formul{\ae} (\ref{cargam}) and (\ref{cargae}) and
integrate over $\rho$ from $\rho=0$ to $\rho = \rho_*$.
Then, one finds, for the magnetic charge
\be
Q_M = 1 - w^2(\rho_*)  <1
\label{cargames}
\ee
which gives a noninteger value, this showing that the solution is
non-topological in the sense discussed in \cite{hosr}-\cite{hos}
for  Einstein-Yang-Mills dyon solutions in asymptotically AdS
space.

Concerning the electric charge, formula
(\ref{cargae}) gives
\be
Q_E = \left.R^2\frac{du}{d\rho} \right|_{\rho = {\rho_*}}
\label{cargaese}
\ee
We represent $Q_E(\bar \rho)$, the charge enclosed in a sphere of
radius $\bar \rho$, for $0\leq\bar \rho < \rho_*$ in figure 7. We
see that as $\bar \rho$ approaches $\rho_*$ the charge grows and
finally diverges. From the numerically determined values for
$\beta$ and $\alpha$, its behavior is given by
\be Q_E({\bar \rho}) \sim (\rho_* - \bar \rho)^{-0.09}.
\ee
%%
%%%%%%%%%%%%%%%%%%%%%%%%%%%%%%%%%%%%%%%%%%%%%%%%%%%%%%%%%%%%%%%%%%%%%%

% Indeed, in order to compute the electric and
%magnetic charges, one should use formul{\ae} (\ref{cargam}) and
%(\ref{cargae}) and integrate over $\rho$ excluding the singular
%point at $\rho = \rho_*$ so that the integration interval becomes
%$(0,\rho_*^-)$.
%%
%Then, one finds, for the magnetic charge
%%
%\be
%Q_M = 1 - w^2(\rho_*^-)  <1
%\label{cargames}
%\ee
%%
%which gives a noninteger value, this showing that the solution is
%non-topological in the sense discussed in \cite{hosr}-\cite{hos}
%for  Einstein-Yang-Mills dyon solutions in asymptotically AdS
%space.
%
%Concerning the electric charge, formula
%(\ref{cargae}) gives
%%
%\be
%Q_E = \left.R^2\frac{du}{d\rho} \right|_{\rho = {\rho_*^-}}
%\label{cargaese}
%\ee
%%
%We represent $Q_{\bar \rho}$, the charge enclosed in a sphere of
%radius $\bar \rho$, for $0\leq\bar \rho < \rho_*$ in figure 7. We
%see that as $\bar \rho$ approaches $\rho_*$ the charge grows and
%from the numerically determined values for $\beta$ and $\alpha$,
%its behavior is given by
%%
%\be
%Q_E({\bar \rho}) \sim (\rho_* - \bar \rho)^{-0.09}
%\ee
% %

~

It is important to stress that although we were not able to fix
the exponents $\alpha$ and $\delta$ analytically, inequalities
(\ref{desi}) allow us to  ensure that all dyon solutions
correspond to bags of gold. Moreover, there is no loss of
generality in our results because we have put $g=1$ from the
beginning: they are valid for arbitrary $g$. Then, in a sense we
can say that the bag of gold singular behavior of dyon solutions
is ``universal''.

%%%%%%%%%%%%%%%%%%%%%%%%%%%%%%%%%%%%%%%%%%%%%%%%%%%%%%%%%%%%%%%
\begin{figure}%[ht]
\centerline{
\psfig{figure=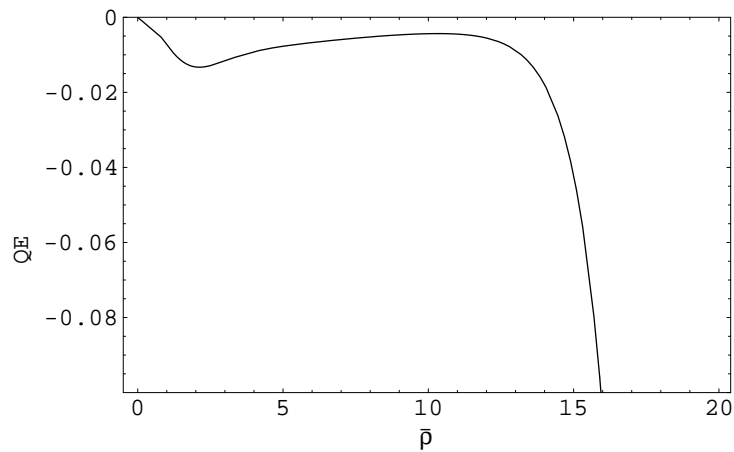,height=6cm,angle=0}}
\smallskip
\caption{ The electric charge  enclosed in a
sphere of radius $\bar \rho$ for the dyon solution
with $w^{(2)} = 0.3  u^{(0)} = -0.01, a^{(0)} = 0.00414 $ \label{fig6} }
\end{figure}
%%%%%%%%%%%%%%%%%%%%%%%%%%%%%%%%%%%%%%%%%%%%%%%%%%%%%%%%%%%%%%%

\section{Summary}
We have discussed in this work classical monopole and dyon
solutions to the equations of motion of $d=4$, ${\cal N} = 4$
Freedman-Schwarz supergravity. Purely magnetic solutions were
already found in \cite{ChV1}-\cite{V3}. Apart from the exact BPS
solution, the authors in reference \cite{V3} constructed,
numerically,  some non-BPS magnetic monopole solutions noting the
existence of singular (bag of gold) solutions in some range of the
shooting parameter. We have confirmed this result, showing
analytically a set of exponents characterizing the singularity.

We have then extended the analysis to allow for electrically
charged dyon solutions. A first important result is that  one can
see that there is no possibility of finding dyon configurations
making the supersymmetry variations to vanish. Then, we conclude
that BPS solutions cannot be electrically charged. Concerning the
second order equations of motion, we have shown analytically that
they all correspond to bag of gold configurations (in contrast
with the purely magnetic case where both regular and singular
solutions were allowed). In fact, we have computed the affine
length ${\cal L}$ along radial null geodesics showing that ${\cal
L}$ is finite and hence $\rho = \rho_*$ corresponds to a real
singularity of the space-time. We have studied the asymptotic
behavior near the singularity finding that the magnetic charge is
not an integer -the solution then corresponds to a non-topological
soliton as those discussed in the case of Einstein-Yang-Mills
theory in asymptotically AdS space \cite{hosr}-\cite{hos}. As
stressed in the introduction, soliton solutions in the bosonic
sector of  $d=4$ supergravity models are of interest, after
uplifting to $d=10$ dimensions, in the context of the AdS/CFT
correspondence. An interesting question in this context is whether
the $d=4$ space-time singularity in $\rho = \rho_*$ survives the
uplifting procedure and, in the affirmative, how it affects the
physical outcome of the $d=10$ theory.

%\vspace{.8 cm}

%\noindent\underline{Acknowledgements}:
\begin{acknowledgements}
We thank H.~Fanchiotti,
P.~Forg\'acs, S.~Reuillon, G.~Ro\-ssi\-ni, M.~Schvellinger,
G.~Silva and M.~Volkov for interesting and useful comments at
different stages of this investigation. This work  was partially
supported by UNLP, CICBA, CONICET, ANPCYT (PICT grant 03-05179)
Argentina and ECOS-Sud Argentina-France collaboration (grant
A01E02).  E.F.M. is partially supported by Fundaci\'on Antorchas,
Argentina.
\end{acknowledgements}

%
%%%%%%%%%%%%%%%%%%%%%%%%%%%%%%%%%%%%%%%%%%%%%%%%%%%%%%%%%%%%%%%%%

\end{document}